\documentclass[reprint,aps,pra,twocolumn,superscriptaddress,10pt,nofootinbib,longbibliography,floatfix]{revtex4-2}
\usepackage[T1]{fontenc}
\usepackage{ORI_Group_style}
\usepackage{physics}
\usepackage{amsmath,amsfonts,bbm, graphicx, color}
\usepackage{amssymb}
\usepackage{appendix}
\usepackage{relsize}
\usepackage{xcolor}
\usepackage{scalerel}
\usepackage{tikz}
\usetikzlibrary{svg.path}

\usepackage[colorlinks]{hyperref}

\hypersetup{
	bookmarksnumbered,
	pdfstartview={FitH},
	citecolor={blue},
	linkcolor={black},
	urlcolor={blue},
	pdfpagemode={UseOutlines}}
\definecolor{darkgreen}{RGB}{20,100,20}
\definecolor{darkblue}{RGB}{0,0,130}
\definecolor{darkred}{rgb}{.8,0,0}

\begin{document}
	
\title{Optimization of Static Potentials  for Large Delocalization and Non-Gaussian Quantum Dynamics of Levitated Nanoparticles Under Decoherence}

\author{Silvia Casulleras}

\author{Piotr T. Grochowski}

 \affiliation{Institute for Quantum Optics and Quantum Information of the Austrian Academy of Sciences, 6020 Innsbruck, Austria}
 \affiliation{Institute for Theoretical Physics, University of Innsbruck, 6020 Innsbruck, Austria}
 
\author{Oriol Romero-Isart}
 \affiliation{Institute for Quantum Optics and Quantum Information of the Austrian Academy of Sciences, 6020 Innsbruck, Austria}
 \affiliation{Institute for Theoretical Physics, University of Innsbruck, 6020 Innsbruck, Austria}
 \affiliation{ICFO - Institut de Ciencies Fotoniques, The Barcelona Institute of Science and Technology, Castelldefels (Barcelona) 08860, Spain}
 \affiliation{ICREA, Passeig Lluis Companys 23, 08010, Barcelona, Spain}

\begin{abstract}
    Levitated nanoparticles provide a  controllable and isolated platform for probing fundamental quantum phenomena at the macroscopic scale.  In this work, we introduce an  optimization method to determine optimal static potentials for the generation of largely delocalized  and non-Gaussian quantum states of levitated nanoparticles. Our optimization strategy accounts for position-dependent  noise sources originating from the fluctuations of the potential.  We provide key figures of merit that allow for fast computation and capture relevant features of the dynamics, mitigating the computational demands associated with the multiscale simulation of this system.
    Specifically, we introduce coherence length and coherent cubicity as signatures of large delocalization and quantum non-Gaussian states, respectively. We apply the optimization approach to a family of quartic potentials and show that the optimal configuration depends on the strength and nature of the noise in the system. Additionally, we benchmark our results with the full quantum dynamics simulations of the system for the optimal potentials.  
\end{abstract}

\maketitle

\section{Introduction}

Levitated nanoparticles offer a controlled experimental platform for investigating quantum phenomena at the interface between classical and quantum mechanics~\cite{Millen_2020,science.abg3027}. Recently,  ground-state cooling of nanoparticles in optical traps was achieved ~\cite{science.aba3993,Magrini2021,Nature10.1038,Kamba:22,PhysRevResearch.4.033051,s41567-023-01956-1,s41467-023-43745-7}. This milestone motivates efforts to observe quantum phenomena at large scales, that is, to prepare non-Gaussian states that are delocalized over scales beyond the zero-point motion, even approaching the size of the particle~\cite{rodallordes2023macroscopic,doi:10.1073/pnas.2306953121}.
 Such states, involving nanoparticles with billions of atoms, could enable quantum matter-wave interferometry with mesoscopic particles~\cite{PhysRevLett.107.020405,ncomms5788,PhysRevLett.117.143003,Stickler_2018,rodallordes2023macroscopic,doi:10.1073/pnas.2306953121} and testing the quantum superposition principle in regimes where collapse models predict the breakdown of quantum mechanics~\cite{PhysRevA.84.052121,RevModPhys.85.471}. Additionally, they could serve as ultraprecise sensors~\cite{PhysRevD.92.062002,Stickler_2018,PhysRevLett.121.063602,PhysRevLett.127.023601} and could allow to explore the interplay between quantum mechanics and gravity, e.g., by measuring the gravitational field of a massive object in a macroscopic quantum superposition~\cite{Rickles,PhysRevD.98.126009}.

\begin{figure}[t]
	\centering
		\includegraphics[width=0.9\linewidth]{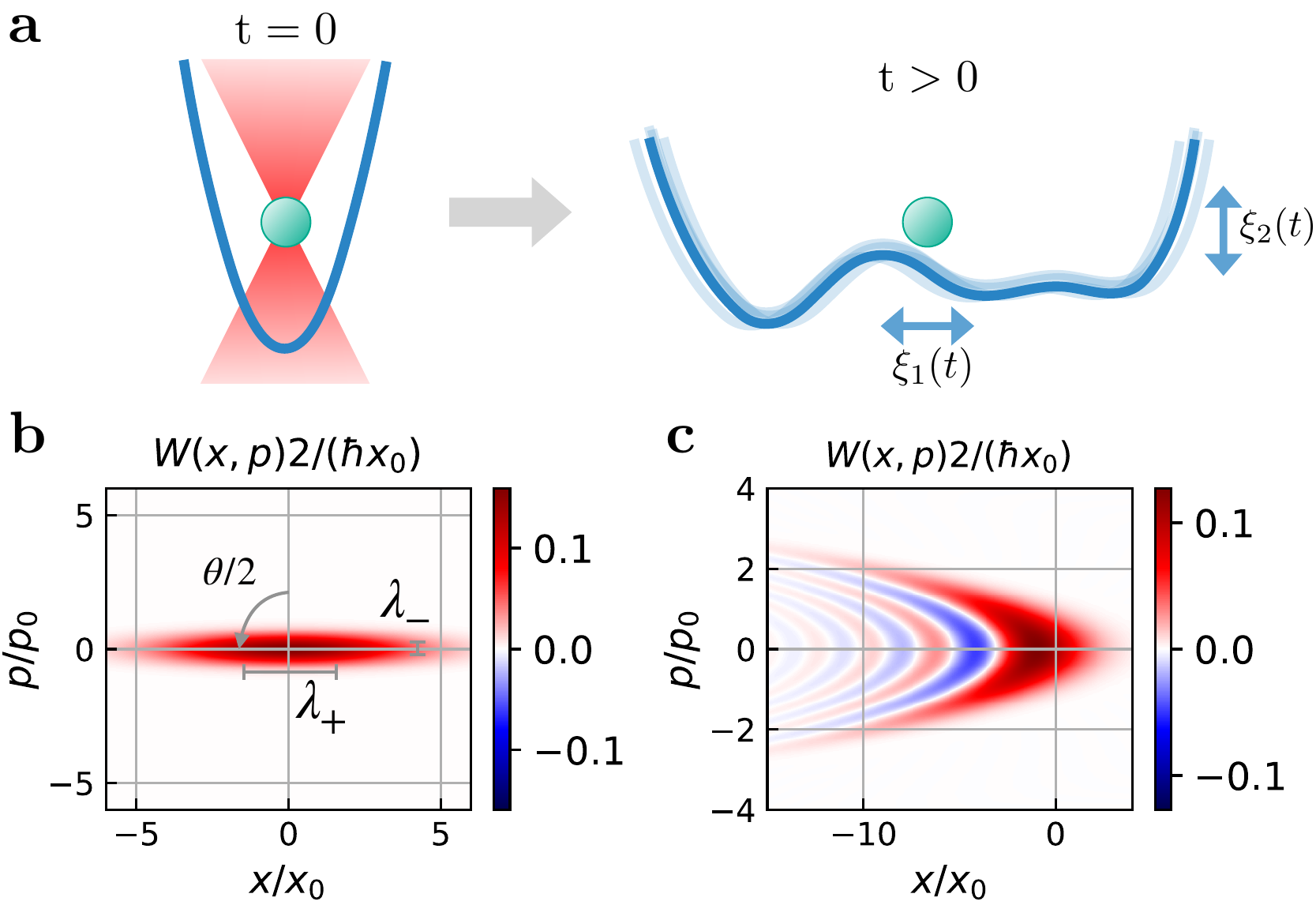}
    \caption{\label{fig1_scheme} a. Sketch of a nanoparticle trapped by an optical tweezer and ground-state cooled at $t=0$. For $t>0$, the particle evolves in a nonharmonic, static, wide, non-optical potential.  The position and amplitude of the potential fluctuate according to the stochastic functions $\xi_1(t)$ and  $\xi_2(t)$, respectively. The dominant dynamics generated by a wide nonharmonic potential correspond to squeezing and generation of cubic-phase states. b. Wigner function of a squeezed state with squeezing parameter $r=1$, squeezing angle $\theta=\pi$, and purity $\mathcal{P}=1$. The standard deviation of the state along the direction of maximum squeezing (expansion) is given by $\lambda_-=e^{-r}$ ($\lambda_+=e^r$). c. Wigner function of a cubic-phase state along the $p$-quadrature with cubicity $\kappa=1$ and purity $\mathcal{P}=1$.
  }
\end{figure}

Efforts towards controlling the motional state of levitated nanoparticles in the quantum regime involve coupling nanoparticles to external nonlinear systems, e.g.,~trapped ions~\cite{bykov2024nanoparticle} or superconducting qubits~\cite{PhysRevLett.109.147205,s41534-020-00333-7}, or internal two-level systems, such as nitrogen-vacancy centers \cite{ncomms12250,mi12060651,PhysRevLett.117.143003,PhysRevResearch.4.023087}. One alternative approach consists in utilizing nonharmonic potentials to generate non-Gaussian dynamics \cite{s41534-021-00453-8,rodallordes2023macroscopic,doi:10.1073/pnas.2306953121}.
 Due to substantial decoherence in optical potentials~\cite{PhysRevLett.116.243601,PhysRevA.108.033714}, it is  beneficial that the dynamics happen either in the absence of a permanent optical potential~\cite{PhysRevLett.107.020405,doi:10.1073/pnas.2306953121,s41534-021-00453-8} or within a nonoptical (``dark'') potential~\cite{Pino_2018,rodallordes2023macroscopic}, e.g.
 generated by electric or magnetic fields~\cite{PhysRevLett.114.123602,nanolett.8b01414, PhysRevApplied.13.064027,PhysRevResearch.3.013018,PhysRevApplied.19.054047,PhysRevLett.131.043603,bonvin2023state}. 
 The mismatch of length scales between the ground-state zero-point motion of the particle and the trapping potential necessitates expanding the wave function, i.e., generating motional squeezing, to explore nonharmonicities and attain larger quantum states~\cite{rosiek2023quadrature,rodallordes2023macroscopic}.
 However, such expansion enhances decoherence effects~\cite{PhysRevLett.127.023601}, making the optimal design of the potential landscape---either through dynamic control or static geometry---crucial for creating sufficiently pure and large quantum states. In this context it is timely and interesting to develop tools to find optimal potential shapes that allow for the preparation of large quantum non-Gaussian states in the presence of decoherence.

In this work, we introduce an optimization approach to obtain optimal wide, static, nonharmonic potentials for the generation of largely delocalized states and non-Gaussian quantum states of levitated nanoparticles.
We perform the optimization in the presence of position-dependent noise stemming from the fluctuations of the position and amplitude of the potential~\cite{PhysRevA.58.3914,10.1007/s003400050823,PhysRevA.59.3766}, in addition to other sources of decoherence such as the emission of thermal photons from the particle~\cite{PhysRevA.84.052121,ncomms5788}.
The static potential is assumed to be generated electrically or magnetically in order to avoid decoherence from photon recoil heating~\cite{PhysRevLett.116.243601,Pino_2018,PhysRevA.108.033714}. Moreover, we focus on potentials that allow for dynamical protocols which are faster than the typical collision time with a gas molecule in ultra-high vacuum~\cite{PhysRevA.84.052121,BF01725541,PhysRevLett.127.023601,doi:10.1073/pnas.2306953121,rodallordes2023macroscopic}, allowing us to neglect the decoherence due to the presence of gas molecules.
The numerical simulation of the system is highly computationally demanding due to the multiscale character of the dynamics \cite{rodallordes2023numerical},  thus hindering the optimization.
To circumvent this problem, we introduce key figures of merit that allow for fast computation and capture signatures of large delocalizations and non-Gaussian quantum states. As an example, we apply our method to find optimal quartic potentials for our figures of merit depending on the noise type and strength.
Additionally, we perform numerically exact simulations of the evolution of the particle for the optimal configurations and show that the figures of merit introduced provide meaningful information about the full quantum dynamics.

This article is structured as follows. First, in Section~\ref{section-dynamics}, we introduce the framework we use to describe the evolution of a levitated nanoparticle in a static potential in the presence of decoherence.
Then, in Section~\ref{section-figs-of-merit}, we define the optimization problem and introduce two figures of merit that capture relevant dynamics of the system, namely the coherence length and the coherent cubicity.
In Section~\ref{section-quartic} we show an example of optimizing quartic potentials for maximizing the above-mentioned figures of merit and benchmark the results with the numerical simulation of the full quantum dynamics of the system.
Finally, in Section~\ref{section-conclusions} we present our conclusions and outlook. 

\section{Dynamics in a wide potential}\label{section-dynamics}

Let us consider a particle of mass $m$ which is optically levitated in a harmonic potential of frequency $\Omega$. Let us assume that the motional state of the particle along one certain direction, labeled by $x$, is prepared in the ground state of the Hamiltonian $\hat{H}_0\equiv\hat{p}^2/(2m)+m\Omega^2 \hat{x}^2/2$. Here, $\hat{x}$ and $\hat{p}$ are the position and momentum operators along the $x$ direction. The zero-point position and momentum fluctuations of the particle are given by $x_0=\sqrt{\hbar/(2m\Omega)}$ and $p_0 = \hbar/(2x_0)$, respectively. At the time $t=0$, the optical potential is suddenly switched off and the particle evolves in a wide static nonharmonic potential $V_\text{s}(x)$ (see Fig.~\ref{fig1_scheme}a).  
The Hamiltonian of the system for $t>0$ is given by 
\be 
\hat{H}_\text{s}\equiv\frac{\hat{p}^{2}}{2m}+V_\text{s}(\hat{x}). \label{hamiltonian_static}
\ee
Our goal is to optimize the shape of the nonharmonic potential $V_\text{s}(x)$ in order to maximize the nonclassical features of the evolution of the state. 
The potential is assumed to be wide, that is, the relevant length scale of the potential is considered to be much larger than the initial spatial extent of the position of the particle. This allows for coherent expansion~\cite{rodallordes2023macroscopic,bonvin2023state} in order to produce a sufficiently delocalized state that is able to explore the weak nonharmonicities of $V_\text{s}(x)$.

In this section, we describe the evolution of a particle within a wide, nonharmonic fluctuating potential $V_\text{s}(x)$, using the framework presented in~\cite{rieracampeny2023wigner}. Following the methodology of~\cite{rieracampeny2023wigner}, we present the master equation that incorporates potential fluctuations and other sources of decoherence. First, we solve the noisy dynamics within the Gaussian approximation. We then describe the non-Gaussian approximation of the coherent dynamics detailed in~\cite{rieracampeny2023wigner}. These two distinct approximations of the particle dynamics will enable us to define the figures of merit introduced in Section III.
 
\subsection{Dynamics in a fluctuating potential}

Let us describe the dynamics of the particle in the centroid frame, obtained by applying the unitary transformation
$\hat{U}_\text{c}(t)\equiv\exp\left[\text{i}\left(\hat{x}p_\text{c}(t)-\hat{p}x_\text{c}(t)\right)/\hbar\right]$~\cite{PhysRevE.64.056602,rieracampeny2023wigner,Bialynicki-Birula_2023}. Here, $x_\text{c}(t)$ and $p_\text{c}(t)$ are the classical trajectories associated to
the Hamiltonian \eqref{hamiltonian_static}, with the initial conditions $x_\text{c}(0)= \langle\hat{x}\rangle(0)$ and $p_\text{c}(0)= \langle\hat{p}\rangle(0)$. Hereafter, we assume $\langle\hat{x}\rangle(0)=\langle\hat{p}\rangle(0)=0.$  The transformation $\hat{U}_\text{c}(t)$ represents a displacement operation to follow the classical trajectories of the state. In the case when the quantum fluctuations of the state around the classical trajectories are small compared to its size,  the transformation to the centroid frame removes the dynamics of the mean position and momentum values of the particle. 

Let us assume that the static potential $V_\text{s}(x)$ fluctuates stochastically in position and amplitude. In that case, the particle experiences an effective time-dependent potential given by $V_\text{f}(x,t)\equiv V_\text{s}(\hat{x}+l\xi_{1}(t))\left[1+\xi_{2}(t)\right]$, where $l$ is a length scale associated to the size of the potential. Here,
$\xi_{1}(t)$ and $\xi_{2}(t)$ represent the stochastic functions that model the fluctuations of the potential in position and amplitude,  respectively. Let us consider that both stochastic functions 
represent a Gaussian white noise and are uncorrelated. 
More specifically, they fulfill the relations $\langle\xi_{i}(t)\rangle=0$ and $\langle\xi_{i}(t)\xi_{j}(t')\rangle=2\pi S_{j}\delta_{ij}\delta(t-t')$ for $j\in\{1,2\}$, where $\langle \cdot \rangle$ denotes the average over many repetitions and $S_j$ is the noise strength. 
 
One can derive an effective master equation for the particle in the centroid frame, accounting for the decoherence due to the fluctuations of the potential as well as other sources of decoherence modelled by a constant displacement noise~\cite{rieracampeny2023wigner}. In particular, the evolution of the particle is given by 
\be
\frac{\partial\hat{\rho}_c}{\partial t}=\frac{1}{i\hbar}\left[\hat{H}_\text{c}(t),\hat{\rho}_{c}(t)\right]+ \mathcal{D}_\text{c}(t)[\hat{\rho}_\text{c}], \label{vonNeumann_classical}
\ee
where $\hat{\rho}_\text{c}=\hat{U}_\text{c}\rho \hat{U}_c^\dagger$ is the state of the particle in the centroid frame, $\hat{H}_\text{c}(t)$ is a time-dependent Hamiltonian and $\mathcal{D}_c(t)$ is a time-dependent decoherence superoperator. The coherent part of the dynamics is given by the  Hamiltonian \eqref{hamiltonian_static} in the centroid frame, that is,
\be
\hat{H}_\text{c}(t)\approx\frac{\hat{p}^{2}}{2m}+m\omega^2\sum_{n=2}^{N}\frac{1}{n}\frac{\alpha_n(t)}{x_0^{n-2}}\hat{x}^n, \label{hamiltonian-cf}
\ee
where we have expanded the nonharmonic potential $V_\text{s}(x)$ around the classical trajectory $x_\text{c}(t)$ up to order $N$. Here, $\omega$ is a frequency scale and we have defined $\alpha_n(t)\equiv V_s^{(n)}(t)x_0^{n-2}/(m\omega^2 (n-1)!)$. Additionally, we have used the short-hand notation
$V_\text{s}^{(n)}(t)\equiv [\partial_x^n V_\text{s}(x)]_{x=x_\text{c}(t)}$. The incoherent part of the dynamics corresponds to the decoherence superoperator~\cite{rieracampeny2023wigner}
\be
\mathcal{D}_\text{c}(t)[\hat{\rho}_\text{c}] \approx -\frac{(\Gamma_\text{f}(t)+\Gamma_0)}{2x_0^2}[\hat{x},[\hat{x},\hat{\rho}_\text{c}]],\label{vonNeumann_dissipation}
\ee
where $\Gamma_\text{f}(t)$ is a time-dependent decoherence rate that models the decoherence due to the fluctuations of the potential, and $\Gamma_0$ is a constant decay rate that models other sources of decoherence, such as thermal emission of the particle.
In particular, $\Gamma_\text{f}(t)$ is given by
\be
\Gamma_\text{f}(t)=\frac{2\pi x_0^2}{\hbar^2}\left(S_1 l^2[V_\text{s}^{(2)}(t)]^2+S_2  [V_\text{s}^{(1)}(t)]^2\right) \label{gamma-fluctuations}.
\ee
Equation \eqref{vonNeumann_dissipation} is obtained assuming that the fluctuations of the potential and the quantum fluctuations of the state around the classical trajectories are small. Note that the latter condition, which corresponds to $|\langle \hat{x}\rangle(t)-x_c(t)|\ll |x_\text{c}(t)|$, is usually satisfied in the scenario that we are interested in, namely a particle evolving in a wide static potential~\cite{rieracampeny2023wigner}. These assumptions allow us to approximate $\mathcal{D}_\text{c}(t)$ by a quadratic superoperator, neglecting the contribution of higher-order commutators in $\hat{x}$.  Note that the decay rate $\Gamma_\text{f}(t)$ is time-dependent due to the different effect that the fluctuations have on the particle depending on its position with respect to the potential.

Assuming that the quantum fluctuations of the state around the classical trajectories are small, one can perform further approximations on the coherent part of the dynamics. First, one can consider the Gaussian approximation, which corresponds to keeping only the lowest-order terms in the operator $\hat{x}$ in the Hamiltonian \eqref{hamiltonian-cf}, namely the quadratic terms. Alternatively, one can also consider the contribution of the first non-Gaussian term to the coherent dynamics, given by the cubic term in equation \eqref{hamiltonian-cf}.  In the following subsections, we describe the evolution of the particle under these two approximations.

\subsection{Gaussian dynamics approximation}\label{sec-gaussian}

Let us describe the first approximation, which corresponds to considering the lowest-order term ($n=2$) in equation  \eqref{hamiltonian-cf}. In this case, the time-dependent Hamiltonian \eqref{hamiltonian-cf} is approximated by the quadratic Hamiltonian
\be
\hat{H}_\text{c}(t)\approx \hat{H}_\text{G}(t)	\equiv\frac{\hat{p}^{2}}{2m}+\frac{1}{2}m\omega^{2}\alpha_2(t)\hat{x}^{2}.\label{eq:gaussian-H}
\ee
Note that, since the initial state of the particle is Gaussian, the quadratic Hamiltonian  $\hat{H}_\text{G}(t)$ generates purely Gaussian dynamics. In addition, the dissipative dynamics described by equation \eqref{vonNeumann_dissipation} are also quadratic. Therefore, the full evolution of the state described by the master equation \eqref{vonNeumann_classical} under the approximation \eqref{eq:gaussian-H} is Gaussian. In that case, the state of the particle can be unequivocally described by the covariance matrix $\mathbf{C}(t)$, whose entries are defined as  $C_{xx}(t)	= \langle\hat{x}^{2}\rangle(t)-\langle\hat{x}\rangle^{2}(t)$, $C_{pp}(t)	= \langle \hat{p}^{2}\rangle(t)-\langle \hat{p}\rangle^{2}(t)$, and $C_{xp}(t)	=C_{px}(t)= \langle\hat{x}\hat{p}+\hat{p}\hat{x}\rangle(t)/2-\langle\hat{x}\rangle(t)\langle\hat{p}\rangle$(t)~\cite{RevModPhys.84.621}. Since $\langle \hat{x}\rangle(0) =\langle \hat{p}\rangle(0) =0$ and the evolution equation contains only quadratic terms, the first-order moments vanish during the whole evolution, \ie, $\langle\hat{x}\rangle(t)=\langle\hat{p}\rangle(t)=0$. Specifically, the equations of motion for the elements of the covariance matrix are given by 
\begin{align}
\partial_tC_{xx}(t)&=\frac{2}{m}C_{xp}(t),\nonumber \\
\partial_tC_{pp}(t)&=-2m\omega^{2}\alpha_2(t)C_{xp}(t)+\frac{\hbar^{2}}{x_0^{2}}(\Gamma_f(t)+\Gamma_0), \nonumber\\
\partial_tC_{xp}(t) &=\frac{1}{m}C_{pp}(t)-m\omega^{2}\alpha_2(t)C_{xx}(t),\label{gaussian-second-moments}
\end{align}	
with the initial conditions $C_{xx}(0)=x_0^2$, $C_{pp}(0)=p_0^2$ and   $C_{xp}(0)=x_0p_0$, where $x_0$ and $p_0$ are the position and momentum zero-point fluctuations.

The evolution of the state under Gaussian dynamics consists of squeezing and rotation in phase space. Thus, the evolution can be described by the squeezing parameter  $r(t)$ and the squeezing angle $\theta(t)$, schematically depicted in Fig.~\ref{fig1_scheme}b.  In particular, the squeezing parameter $r(t)$ is given by $r(t)=\ln(\lambda_-(t))$, where $\lambda_-(t)$ is the smallest eigenvalue of the dimensionless covariance matrix $\mathbf{\bar C}(t)$~\cite{Idel_2016}. The entries of $\mathbf{\bar C}(t)$ are defined as $\bar{C}_{xx}(t)=C_{xx}(t)/x_0^2$, $\bar{C}_{xp}(t)=\bar{C}_{px}(t)=C_{xp}(t)/(x_0p_0)$ and $\bar{C}_{pp}(t)=C_{pp}(t)/p_0^2$. The squeezing angle $\theta(t)$  corresponds to 
\be
\theta(t)=2\arctan\left(\frac{\mathbf{e}_p\cdot \mathbf{u}(t)}{\mathbf{e}_x\cdot \mathbf{u}(t)}\right),\label{squeezing-angle}
\ee
where $\mathbf{u}(t)$ is the eigenvector of $\mathbf{\bar{C}}(t)$ associated to the eigenvalue $\lambda_-(t)$. Here, $\mathbf{e}_x$ and $\mathbf{e}_p$ denote the unit vectors along the direction $x$ and $p$, respectively. The purity of the state can also be calculated from the dimensionless covariance matrix, as 
\be
\mathcal{P}(t)=\frac{1}{\sqrt{\det\left[\bar{ \mathbf{C}}(t)\right]}}.\label{purity}
\ee

\subsection{Approximate non-Gaussian dynamics evolution}\label{subsec-non-Gaussian}

Let us describe a second approximation, which extends beyond the one introduced in Section \ref{sec-gaussian}. Here, we consider both the quadratic and cubic terms in equation \eqref{hamiltonian-cf}, that is
 \be
\hat{H}_\text{c}(t)\approx \hat{H}_\text{nG}(t)\equiv\hat{H}_\text{G}(t)+\frac{m\omega^2}{3x_0}\alpha_3(t)\hat{x}^3, \label{potential-nG}
\ee
where $\hat{H}_\text{G}(t)$ is the quadratic Hamiltonian given by equation \eqref{eq:gaussian-H}. Note that the Hamiltonian $\hat{H}_\text{nG}(t)$ corresponds to the lowest-order terms in equation \eqref{hamiltonian-cf} leading to an evolution of the particle beyond Gaussian dynamics. 

In order to simplify the description of the dynamics generated by the nonharmonic Hamiltonian \eqref{potential-nG}, let us first apply a transformation to the Gaussian frame driven by the Hamiltonian $\hat{H}_\text{G}(t)$~\cite{rieracampeny2023wigner}. Specifically, the frame transformation is defined by the unitary transformation $\hat{U}_\text{G}(t)\equiv\exp_+\left[\text{i}\int_0^t \text{d}t'\hat{H}_\text{G}(t')/\hbar\right]$, where $\exp_+$ denotes the time-ordered exponential and $\hat{H}_\text{G}(t)$ is given by equation \eqref{eq:gaussian-H}. The position and momentum operators in the Gaussian frame, given by  $\tilde{\hat{x}}=\hat{U}_\text{G}^\dagger(t)\hat{x}\hat{U}_\text{G}(t)$ and  $\tilde{\hat{p}}=\hat{U}_\text{G}^\dagger(t)\hat{p}\hat{U}_\text{G}(t)$, respectively, can be computed as $(\tilde{\hat{x}}/x_0,\tilde{\hat{p}}/p_0)^\top=\mathbf{S}(t)(\hat{x}/x_0,{p}/p_0)^\top$, where $\mathbf{S}(t)$ is a symplectic matrix that can be obtained as the solution to the differential equation~\cite{rieracampeny2023wigner},
\be
\frac{\partial\mathbf{S}(t)}{\partial t}=\left(\begin{array}{cc}
0 & \Omega\\
-\omega^{2}\alpha(t)/\Omega & 0
\end{array}\right)\mathbf{S}(t).
\ee
Let us write the position operator in the Gaussian frame as
$\hat{\tilde{x}}=\eta(t)(\cos\phi(t)\hat{x}+\sin\phi(t)\hat{p}x_0/p_0)$, where $\eta(t)$ and $\phi(t)$ can be calculated from the elements of the matrix $\mathbf{S}(t)$ as $\eta(t)=\sqrt{S_{xx}^2(t)+S_{xp}^2(t)}$, and $\phi(t)=\arctan{(S_{xp}(t)/S_{xx}(t))}$. Note that $\eta(t)$ and $\phi(t)$ can be interpreted as the quadrature squeezing and quadrature angle that define the transformation to the Gaussian frame, respectively. The Hamiltonian \eqref{potential-nG} in the interaction  frame driven by the Gaussian dynamics is then given by
\be
\hat{\tilde{H}}_\text{nG}(t)=\frac{2m\omega\Omega\beta(t)}{x_0}\left[ \cos(\phi(t))\hat{x}+\sin(\phi(t)) \frac{x_0}{p_0}\hat{p}\right]^3, \label{potentian-nG}
\ee
where we have defined $\beta(t)\equiv 3\alpha_3(t)\eta^3(t)/2$. Equation \eqref{potentian-nG} shows that the effect of the non-Gaussian dynamics is enhanced by a large quadrature squeezing $\eta(t)$. The evolution operator associated with the non-Gaussian dynamics \eqref{potentian-nG} is given by
\be
\hat{U}_\text{nG}(t)=\exp_+\left[-\text{i}\int_0^t\text{d}t' \omega \beta(t')\frac{\hat{x}_{\phi(t)}^3}{x_0^3}\right],\label{evolution-nG}
\ee
where $\hat{x}_{\phi(t)}$ denotes the rotated quadrature $\hat{x}_{\phi(t)}\equiv \hat{x}\cos(\phi(t))+\hat{p}\sin(\phi(t)) x_0/p_0$. Solving the dynamics generated by equation~\eqref{evolution-nG} for a general time-dependent quadrature angle $\phi(t)$ involves computationally expensive numerical integration. Thus, instead of solving the full dynamics described by the evolution operator $\hat{U}_\text{nG}(t)$, in Section~\ref{section-figs-of-merit} we will define a figure of merit inspired by equation~\eqref{evolution-nG} that requires only the computation of the functions $\beta(t)$ and $\phi(t)$, which can be obtained from the Gaussian dynamics simulation of the particle.

\section{Potential optimization}\label{section-figs-of-merit}
Our optimization approach consists of finding the optimal potential shape $V_\text{s}(x)$ in equation~\eqref{hamiltonian_static} which maximizes the nonclassical features of the evolution of a particle following equation~\eqref{vonNeumann_classical}. The optimization using a figure of merit that requires a complete simulation of the dynamics of this multiscale problem is excessively computationally challenging~\cite{rodallordes2023numerical}.  Hence, here we introduce two figures of merit that allow for fast computation and capture the main features of the  Gaussian and approximate quantum non-Gaussian evolution of the particle.

\subsection{Coherence length}\label{subsec-coherence}

Let us introduce our first figure of merit, namely the coherence length, which is relevant to describe the Gaussian evolution of the state of the particle. 
We define the coherence length $\xi(t)$ along the direction of maximum expansion of the state as
\be
\xi(t)\equiv\sqrt{8 \lambda_+(t)} \mathcal{P}(t)x_0\label{coherence-length}.
\ee
where $\lambda_+(t)$ denotes the largest eigenvalue of the dimensionless covariance matrix $\mathbf{\bar C}(t)$, and $\mathcal{P}(t)$ is the purity of the Gaussian state, given by equation \eqref{purity}.  It can be shown that $\lambda_+(t)$ corresponds to the variance of the position of the particle along the direction of maximum expansion, that is, $\lambda_+(t)=\langle \hat{x}^2_{\theta(t)}\rangle/x_0^2$. Here,  $\hat{x}_{\theta(t)}$ denotes the rotated quadrature $\hat{x}_{\theta(t)}\equiv \hat{x}\cos(\theta(t))+\hat{p}\sin(\theta(t)) x_0/p_0$, where $\theta(t)$ is the squeezing angle  given by equation \eqref{squeezing-angle}.

The coherence length $\xi(t)$ fulfills the following relation,
\be
\langle-\frac{x_{\theta(t)}}{2}|\hat{\rho}_c|\frac{x_{\theta(t)}}{2}\rangle(t)=\frac{1}{\sqrt{2\pi  \lambda_+(t)}}\exp\left(-\frac{x_{\theta(t)}^2}{\xi^{2}(t)}\right),\label{coh-length-rel}
\ee
where $|x_{\theta(t)}/2\rangle$ denotes the eigenstate of the operator $\hat{x}_{\theta(t)}/2$ with eigenvalue $x_{\theta(t)}/2$. Therefore, the coherence length  provides a measure of the decay of the correlations of the Gaussian state along the direction of maximum expansion. This figure of merit can be useful in scenarios where it is desirable to obtain large expansions while keeping the coherence of the state.

\subsection{Coherent cubicity}\label{subsec-cubicity}

Let us now provide a figure of merit that contains information about the non-Gaussian dynamics given by the nonharmonic potential \eqref{potential-nG}, as well as the coherence of the state, which we denote by coherent cubicity. This figure of merit is inspired by equation~\eqref{evolution-nG}, but does not require the simulation of the non-Gaussian dynamics of the system. 
To this end, let us first introduce a time-dependent coefficient, denoted by cubicity, that quantifies the strength of the non-Gaussian dynamics described by the evolution operator \eqref{evolution-nG}. More specifically, we define the cubicity  $\kappa(t)$ as
\begin{align}
\kappa(t)&\equiv \sqrt{\beta_\text{s}^2(t) +
\beta_\text{c}^2(t)}, \label{kappa2}
\end{align}
where $\beta_\text{s}(t)\equiv\int_0^t \text{d}t'\omega\beta(t')\sin(\phi(t'))$, 
$\beta_\text{c}(t)\equiv\int_0^t \text{d}t'\omega\beta(t')\cos(\phi(t'))$, and the time-dependent functions $\beta(t)$ and $\phi(t)$ are introduced in Section~\ref{subsec-non-Gaussian}. 

Let us provide a motivation for the definition  \eqref{kappa2}. 
During the evolution of a particle in a wide potential, where one obtains large expansions, the quadrature angle $\phi(t)$ is mostly constant, that is, $\phi(t)\approx\phi_0$~\cite{rieracampeny2023wigner}. In that case, the cubicity \eqref{kappa2} reads 
$\kappa(t)\approx \int_0^t \text{d}t'\omega\beta(t'),$ and the evolution operator in equation \eqref{evolution-nG} can be approximated by $\hat{U}_\text{nG}(t)\approx\exp\left[-i\kappa(t)\hat{x}_{\phi_0}^3/x_0^3\right]$~\cite{rieracampeny2023wigner}. Therefore, the dynamics in the case of a constant quadrature angle consists of the generation of a cubic-phase state~\cite{rodallordes2023macroscopic,doi:10.1073/pnas.2306953121,PhysRevA.100.013831,s42005-022-00910-6,Kala:22} along the rotated quadrature $\hat{x}_{\phi_0}= \hat{x}\cos(\phi_0)+\hat{p}\sin(\phi_0) x_0/p_0$ (see Fig.~\ref{fig1_scheme}c for an example of a cubic-phase state). In that case, the cubicity $\kappa(t)$ quantifies the strength of the generator $\hat{U}_\text{nG}(t)$. The definition \eqref{kappa2} of $\kappa(t)$ has been chosen to describe the strength of the generation of cubic-phase states along a general quadrature, allowing for the quadrature angle $\phi(t)$ to be a time-dependent function which is piecewise constant at the times when  $\beta(t)$ is relevant. Specifically, for wide potentials, the possible values of $\phi(t)$ are usually separated by an integer multiple of $\pi$, a property that can be understood in terms of the Maslov index in the semi-classical approximation~\cite{Maslov}. In this way, $\kappa(t)$  allows to account for the accumulation as well as for the decrease of cubicity of the state along a general quadrature.

We are interested in maximizing the effect of the non-Gaussian dynamics while keeping the particle in a coherent state. Hence, we define the coherent cubicity as the product of the absolute value of the cubicity $\kappa(t)$ and the Gaussian purity of the state, that is,
\be
K(t)\equiv |\kappa(t)|\mathcal{P}(t),\label{coherent-cubicity}
\ee
where $\mathcal{P}(t)$ is given by equation \eqref{purity}. Note that we choose Gaussian purity instead of the purity of the full state in order to obtain a figure of merit that does not require the simulation of the non-Gaussian evolution \eqref{evolution-nG}. We also remark that alternative figures of merit could be defined, e.g.,  combinations of different powers of cubicity and purity. However, as we show in the example in Section~\ref{section-quartic}, the coherent cubicity given by equation \eqref{coherent-cubicity} provides valuable insights into the full non-Gaussian evolution of the particle.

\subsection{Constraints on the potential}\label{constraints}

When optimizing the shape of the static potential $V_\text{s}(x)$ for the figures of merit introduced above, we impose several constraints on the potential to meet some requirements regarding experimental feasibility. First, we restrict the set of candidates to potentials that, with respect to the initial position of the particle, lead to closed phase-space classical trajectories.  This condition allows for many repetitions of the protocol while using one single particle. Second, we  set a bound for the derivatives of the potential at the mean position of the particle during one full classical trajectory. In particular, we set $|\alpha_2(t)|\leq \alpha_\text{b}$ for $0<t<T_\text{c}$, where $\alpha_\text{b}$ is a constant value and $T_\text{c}$ is the period of the classical trajectory. Third and finally, we impose that the period of the classical trajectory is smaller than the typical collision time $t_\text{gas}$ with a gas molecule in ultra-high vacuum~\cite{PhysRevA.84.052121,BF01725541,PhysRevLett.127.023601,doi:10.1073/pnas.2306953121,PhysRevLett.132.133602}. This condition
 allows us to neglect the effect of decoherence due to gas molecules during each run of the protocol. The time scale associated with a single gas
scattering event and for a spherical particle of radius $R$
is given by $t_\text{gas} = 3\sqrt{m_\text{gas}k_BT_\text{gas}}/(16\pi \sqrt{2\pi}P_\text{gas}R^2)$ \cite{PhysRevA.84.052121},
where $m_\text{gas}$, $T_\text{gas}$, and $P_\text{gas}$ are the single molecule mass, temperature, and pressure of the gas, respectively. An appropriate condition that complies with current levitated nanoparticle experiments is given by $t_\text{gas}\Omega/(2\pi)\approx 1500$.

\subsection{Optimization algorithm}\label{section-optimization}

We propose to  perform the optimization  of the potential $V_\text{s}(x)$ using a trust-region interior point method~\cite{doi:10.1137/S1052623497325107}. This method combines trust-region and interior point methods to  solve nonlinear optimization problems with constraints. The trust-region component approximates the objective function within a specified region around the current solution. Additionally, the interior point method ensures that the solution remains feasible by using barrier functions that prevent boundary violations. The constraints imposed on the potential, namely $T_\text{c}\leq t_\text{gas}$ and $|\alpha_2(t)|\leq \alpha_\text{b}$, correspond to nonlinear inequality constraints. In particular, $T_\text{c}$ and $\alpha_2(t)$ are computed from the numerical integration of the Hamilton equations. The optimal potential can then be obtained after performing $M$ runs of the optimization algorithm using different randomized initial seeds for the optimization parameters that describe the potential, where $M$ is sufficiently large such that the optimization converges. 

\begin{figure}
	\centering
		\includegraphics[width=1\linewidth]{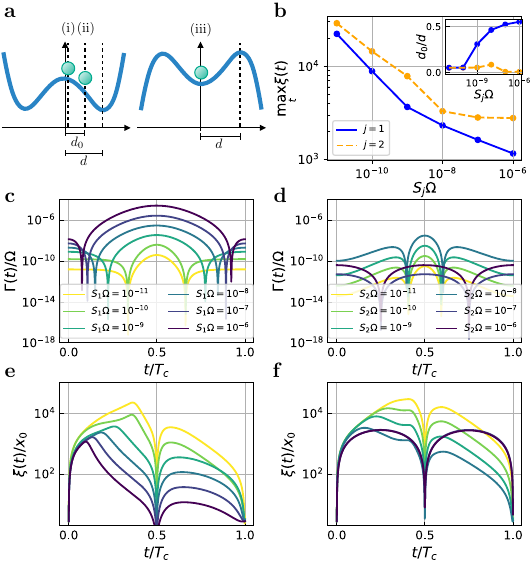}
    \caption{\label{fig2_coh_length}  a. Optimal configurations for maximizing coherence length during one particle trajectory for low potential fluctuations (i), for large values of position fluctuations (ii), and large values of amplitude fluctuations (iii). b. Maximum value of coherence length during one classical trajectory for the optimal potential and optimal $d_0$ (inset) as a function of position fluctuations strength $S_1$ (solid line) and amplitude fluctuations strength $S_2$ (dashed line).  c. Decoherence rate as a function of time associated with position fluctuations of the potential, for different values $S_1$.  d. Decoherence rate associated to amplitude fluctuations for different values of $S_2$. e. Coherence length as a function of time for different position noise strengths $S_1$. f. Coherence length as a function of time for different amplitude noise strengths $S_2$.  Subfigures e and f share the same legend as c and d, respectively. Optimization performed using the ‘trust-constr’ method implemented in SciPy, with the following parameters: $\omega/\Omega=10^{-3}$, $d/x_0=10^6$, $t_\text{gas}\Omega/(2\pi)=1500$, $\alpha_\text{b}=5$,  $M=2000$ randomized initial seeds within the range  $d_0/d\in [0.005,\sqrt{2}]$,  tolerance for termination of $10^{-6}$ and $10^{3}$ maximum number of algorithm iterations. 
  }
\end{figure}

\section{Example: Optimal quartic potentials}\label{section-quartic}
In this section, we focus on applying the optimization method to a particular family of potentials.  Specifically, we consider the family of quartic potentials
\begin{align}
V_\text{q}(x) &= \frac{1}{2}m\omega^{2}\left(a(x-d_{0})^{2}+\frac{b}{2d^{2}}(x-d_{0})^{4}\right)  \label{quartic-family}
\end{align}
where $a$, $b$ and $d_0$ are the optimization parameters. Here,
$\omega$ is a frequency scale, $d$ is a length scale, $d_0$ is the center of the potential, and $a$, $b$ are constants fulfilling $|a|=|b|=1$. The family of potentials \eqref{quartic-family} includes the double-well ($a=-1$, $b=1$) and inverted double-well ($a=1$, $b=-1$) potentials (see Fig.~\ref{fig2_coh_length}a), as well as purely positive ($a=b=1$) and negative ($a=b=-1$) frequency nonharmonic potentials. In particular, the double-well and inverted double-well potentials are good candidates for the generation of nonclassical states since they lead to an expansion of the particle state at the inverted part of the potential, enhancing the nonharmonicity of the potential~\cite{rodallordes2023macroscopic,rieracampeny2023wigner,rosiek2023quadrature}. Our goal is to find the optimal potential within the family \eqref{quartic-family} that maximizes the figures of merit introduced in Sections~\ref{subsec-coherence} and \ref{subsec-cubicity} at any instance of time during one classical trajectory, subject to the constraints introduced in Section~\ref{constraints}. Afterward, we compare the figures of merit for the optimal potentials with the full numerical simulations of the system \cite{rodallordes2023numerical}.  The optimization is performed for different levels and types of noise stemming from the fluctuations of the potential, described by the noise strengths $S_1$ and $S_2$ in equation \eqref{gamma-fluctuations}, where we have set $l=d$. For simplicity, we set $\Gamma_0=0$, that is, we neglect the position-independent sources of decoherence. Here, we fix the frequency scale to $\omega/\Omega=10^{-3}$ and the length scale to $d/x_0=10^{-6}$, values that could be accessible in  experimental implementations of  wide potentials.

\subsection{Optimal potentials for maximum coherence length}
The optimal quartic potential within the family \eqref{quartic-family} that maximizes the figure of merit $\max_{t\in [0,T_\text{c}]}\xi(t)$ depends on the values of the noise strengths $S_1$ and $S_2$. The maximum coherence length achieved by the optimal protocol as a function of noise strength is shown in Fig.~\ref{fig2_coh_length}b. 
Specifically, the optimization leads to three different regimes, corresponding to the double-well (DW) or inverted double-well (IDW) potentials centered at different positions with respect to the initial particle location (Fig.~\ref{fig2_coh_length}a).  

The first regime (i) corresponds to the DW potential ($a=-1$, $b=1$) centered at the initial position of the particle (Fig.~\ref{fig2_coh_length}a). This configuration is optimal for low levels of position and amplitude fluctuations  ($S_1\Omega\lesssim10^{-10}$ and $S_2\Omega\lesssim10^{-9}$). In particular,  the optimal position $d_0$ is the closest to $\langle\hat{x}\rangle=0$ allowed by the condition that the period of the classical trajectory fulfills the constraint $T_\text{c}\leq t_\text{gas}$.  In this configuration, the motional expansion of the state is maximized. The time-dependent decoherence rate $\Gamma_\text{f}(t)$ associated with the fluctuations of the optimal potential is relatively small ($\Gamma_\text{f}(t)/\Omega<10^{-8}$ for $0<t<T_\text{c}$), as shown in Figs.~\ref{fig2_coh_length}c-d. Note that the maximum value of the coherence length is achieved shortly before one half of the period of the classical trajectory (Figs.~\ref{fig2_coh_length}e-f).

\begin{figure}
	\centering
		\includegraphics[width=1\linewidth]{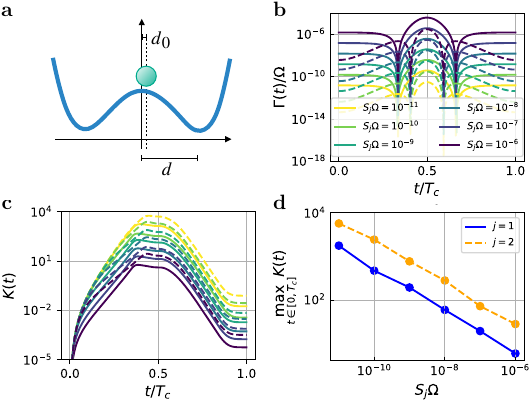}
    \caption{\label{fig3_cubicity} a. Optimal potential for maximizing coherent cubicity during one particle trajectory. The optimal parameters are given by $a=-1$, $b=1$, and $d_0\ll d$ such that $T_\text{c}=t_\text{gas}$. b. Decoherence rate as a function of time associated with position (solid lines) and amplitude (dashed lines) fluctuations of the potential, for different magnitudes $S_j$, where $j=1,2$. c. Coherent cubicity as a function of time for different magnitudes of position (solid) and amplitude (dashed) potential fluctuations. d. The maximum value of the coherent cubicity in the range $t\in[0,T_\text{c}]$ as a function of noise strength $S_1$ (solid line) and $S_2$ (dashed line). Subfigures b and c share the same legend.  Optimization performed using the ‘trust-constr’ method implemented in SciPy, with the following parameters: $\omega/\Omega=10^{-3}$, $d/x_0=10^6$, $t_\text{gas}\Omega/(2\pi)=1500$, $\alpha_\text{b}=5$,  $M=2000$ randomized initial seeds within the range  $d_0/d\in [0.005,\sqrt{2}]$,  tolerance for termination of $10^{-6}$ and $10^{3}$ maximum number of algorithm iterations. 
  }
\end{figure}

The second optimal regime (ii) corresponds to the DW potential centered at the position $d_0=d/2$, where $d$ is the distance from the center to the minima of the double well (Fig.~\ref{fig2_coh_length}a). This configuration is optimal for large levels of noise associated with the fluctuations in the position of the potential ($S_1\Omega\gtrsim 10^{-7}$), and it corresponds to the particle evolving in the region of positive frequency of the DW potential. In this case, the expansion of the maximum variance of the state is smaller, but the purity of the state is preserved compared to a strongly squeezed state, leading to a larger value of coherence length. This configuration minimizes the effect of the potential fluctuations at short times, since the term $V_\text{s}^{(2)}(t)$ in equation \eqref{gamma-fluctuations} vanishes, leading to a maximum coherence length at a time $t\ll T_\text{c}$ (see Fig.~\ref{fig2_coh_length}e).  For intermediate levels of noise ($10^{-9}\lesssim S_1\Omega\lesssim 10^{-8}$), the  optimal configuration is given by a continuous transition between the aforementioned regimes (i)-(ii) (inset of Fig.~\ref{fig2_coh_length}b).

The third optimal regime (iii) corresponds to the IDW potential ($a=1$, $b=-1$), where the particle is initially located at the minimum of the potential (Fig.~\ref{fig2_coh_length}a). The IDW potential is optimal 
for large levels of noise associated with the potential amplitude fluctuations ($S_2\Omega\gtrsim 10^{-7}$).  In this case, the particle experiences an expansion equivalent to the evolution in free space (see Fig.~\ref{fig2_coh_length}f).   This scenario minimizes the decoherence rate $\Gamma_\text{f}(t)$, since $V_\text{s}^{(1)}(t)$ in equation \eqref{gamma-fluctuations} vanishes. For  intermediate levels of amplitude noise ($S_2\Omega\approx10^{-8}$), the optimal potential is the DW slightly displaced from the initial position of the particle (inset of Fig.~\ref{fig2_coh_length}b).

\subsection{Optimal potential for maximum coherent cubicity}
Let us now focus on obtaining the optimal quartic potential within the family \eqref{quartic-family} for maximizing the coherent cubicity \eqref{coherent-cubicity} during one classical trajectory, that is, using the reward function $\max_{t\in [0,T_\text{c}]}K(t)$. The optimization leads to one optimal configuration for all types and levels of noise associated with the potential fluctuations within the studied range ($10^{-11}\leq S_j\Omega\leq 10^{-6}$ for $j=1,2$). The optimal potential is the double-well potential ($a=-1$, $b=1$), with the particle starting at the inverted part of the potential \cite{rodallordes2023macroscopic}, as depicted in Fig.~\ref{fig3_cubicity}a. More specifically, the optimal value of $d_0$ corresponds to the position associated to  the maximum allowed evolution time, i.e., the classical period $T_\text{c}=t_\text{gas}$. This configuration maximizes the expansion of the state of the particle, which is needed to enhance the non-Gaussian dynamics that generate quantum features. In particular, the cubicity $\kappa(t)$ in the DW potential is maximum at $t=T_\text{c}/2$, that is, when the particle evolves in the quartic wall of the potential \cite{rieracampeny2023wigner}. Since the decoherence rate $\Gamma_\text{f}(t)$ associated to the different values of noise $S_1$ and $S_2$ is also maximum at $t=T_\text{c}/2$ (Fig.~\ref{fig3_cubicity}b), the coherent cubicity $K(t)$ is maximum at at time $t\lesssim T_\text{c}/2$, as shown in Fig.~\ref{fig3_cubicity}c. The maximum value of coherent cubicity achieved by the optimal potential decays exponentially with the noise strengths $S_1$ and $S_2$ (see Fig.~\ref{fig3_cubicity}d).

\subsection{Comparison with full quantum dynamics}

\begin{figure}
	\centering
		\includegraphics[width=1\linewidth]{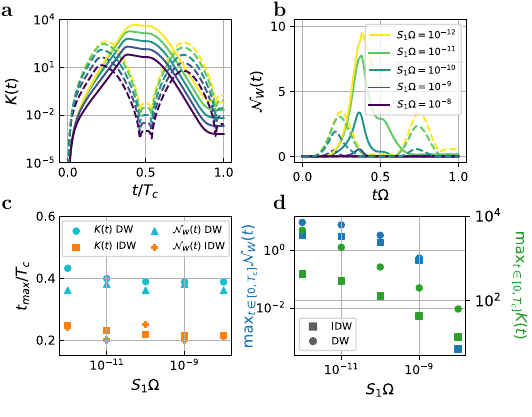}
    \caption{\label{fig4_comparison}  a. Coherent cubicity of a particle evolving in a DW potential (solid lines) and in an IDW potential (dashed lines) as a function of time and noise strength $S_1$. b. Wigner negativity of particle evolving in a DW potential (solid lines) and in an IDW potential (dashed lines) as a function of time and noise strength $S_1$. c. Time $t_\text{max}$ at which the coherent cubicity and the Wigner negativity are maximum for the DW and IDW potentials as a function of noise strength $S_1$. d. Maximum value of the Wigner negativity (left axis) and the coherent cubicity (right axis) as a function of $S_1$ for the DW and IDW configurations. The parameters of the potentials are given by $a=-1$, $b=1$, $d_0/d=0.05$ (DW) and $a=1$, $b=-1$, $d_0/d=0.95$ (IDW), where $d/x_0=10^{6}$ and $\omega/\Omega=10^{-3}$. Subfigures a and b share the same legend.
  }
\end{figure}
Our optimization approach utilizes the figures of merit introduced in Section~\ref{section-figs-of-merit} as signatures of large delocalization and generation of non-Gaussian quantum states. However, these figures of merit are introduced following approximations that do not account for the full quantum dynamics of the system. Hence, in this section, we perform a comparative analysis between our figure of merit that accounts for the leading non-Gaussian term, namely the coherent cubicity, and the Wigner negativity of the state of a particle evolving in potentials within the family \eqref{quartic-family} considered in the optimization approach. The Wigner negativity is calculated from numerical simulations of the full quantum dynamics described by equation \eqref{vonNeumann_classical}~\cite{rodallordes2023numerical}. 

In particular, let us consider two potentials within the family \eqref{quartic-family}, namely the DW and the IDW potentials. As mentioned above, the optimal potential to maximize coherent cubicity is the DW potential. For simplicity, our analysis is centered on one particular type of decoherence, namely the noise stemming from the fluctuations in the position of the potential ($S_2=0$). The behaviour of the coherent cubicity as a function of time for different levels of noise strength (Fig.~\ref{fig4_comparison}a) qualitatively agrees with the behaviour of the Wigner negativity volume (Fig.~\ref{fig4_comparison}b), defined as $\mathcal{N}_\text{W}(t)\equiv \int_\mathbb{R} \text{d}x\text{d}p \left(\left| W(x,p,t)\right|\right)-1$, where $W(x,p,t)$ denotes the Wigner function of the state~\cite{PhysRevA.98.052350}. More specifically, the times at which coherent cubicity and Wigner negativity are maximal coincide for both potentials (see Fig.~\ref{fig4_comparison}c).  Furthermore, the observed trend in the figure of merit as a function of noise strength agrees with the behavior exhibited by the Wigner negativity (Fig.~\ref{fig4_comparison}d). Specifically, the DW potential shows a greater amount of Wigner negativity in comparison to the IDW  for all levels of noise, as predicted by our optimization method. Our findings demonstrate that the proposed figure of merit effectively captures qualitative aspects of the Wigner negativity, thus indicating its potential utility as a metric for assessing the quantum non-Gaussian behavior of a particle evolving in a nonharmonic static potential. 

\section{Conclusions}\label{section-conclusions}

In this work, we have introduced an optimization method aimed at determining optimal static potential shapes for generating largely delocalized states and non-Gaussian quantum states of levitated nanoparticles. This method accounts for position-dependent noise sources inherent to experimental setups. To minimize recoil heating and avoid decoherence due to collisions with gas molecules, we focused on non-optical static potentials and rapid protocols conducted in ultra-high vacuum environments. We considered stochastic fluctuations in the position and amplitude of the potential as primary sources of decoherence in particle dynamics, alongside other decoherence sources. To mitigate the computational demand associated with the multiscale simulation of the system, we proposed key figures of merit  that enable efficient computation while capturing essential dynamic features. Specifically, we introduced coherence length as an indicative measure of the emergence of large delocalization and  coherent cubicity as a signature of the generation of non-Gaussian quantum states.

We have applied our optimization approach to a family of quartic potentials, showing that the optimal configuration depends on the nature and strength of the noise. We found that the optimal quartic potential for maximizing coherence length is either the double-well or the inverted double-well potential, depending on the type of noise, with different positions relative to the initial position of the nanoparticle for varying noise strengths. Additionally, we determined that the optimal potential within the considered family for the generation
of quantum non-Gaussian features, captured by the coherent cubicity, is the double-well potential introduced in \cite{rodallordes2023macroscopic}.  An interesting outlook for our work is to extend the optimization approach to a broader family of potential shapes, including higher-order or time-dependent shaping of potential landscapes~\cite{PhysRevLett.127.023601,grochowski2023quantum}. Our optimization method could be used to explore the possibility to obtain protocols for the generation of macroscopic quantum superpositions of levitated nanoparticles that are robust to environmental decoherence. Moreover, similar optimization methodologies could be employed to develop protocols for the optimal detection of noise sources in non-optical potentials.

\section*{Acknowledgements}
We acknowledge the involvement of K. Kustura in the early stages of this research project and discussions with A. Riera-Campeny, M. Roda-Llordes, O. Benz, and the Q-Xtreme synergy group. This research has been supported by the European Research Council (ERC) under the Grant Agreement No. [951234] (Q-Xtreme ERC-2020-SyG) and by the European Union’s Horizon 2020 research and innovation program under Grant Agreement No. [863132] (IQLev).  The computational results presented here have been achieved in part using the LEO HPC infrastructure of the University of Innsbruck.

\bibliography{Mybib.bib}

\end{document}